\documentclass[12pt]{article}

\usepackage{epstopdf}

\usepackage{natbib}
\bibpunct[, ]{(}{)}{;}{a}{}{,}

\usepackage{color}
\usepackage{fancyvrb}

\DefineVerbatimEnvironment{Highlighting}{Verbatim}{commandchars=\\\{\}}
\usepackage{framed}
\definecolor{shadecolor}{RGB}{248,248,248}
\newenvironment{Shaded}{\begin{snugshade}}{\end{snugshade}}

\newcommand{\CommentTok}[1]{\textcolor[rgb]{0.56,0.35,0.01}{\textit{#1}}}

\newcommand{\DataTypeTok}[1]{\textcolor[rgb]{0.13,0.29,0.53}{#1}}
\newcommand{\DecValTok}[1]{\textcolor[rgb]{0.00,0.00,0.81}{#1}}

\newcommand{\FloatTok}[1]{\textcolor[rgb]{0.00,0.00,0.81}{#1}}

\newcommand{\KeywordTok}[1]{\textcolor[rgb]{0.13,0.29,0.53}{\textbf{#1}}}
\newcommand{\NormalTok}[1]{#1}
\newcommand{\OperatorTok}[1]{\textcolor[rgb]{0.81,0.36,0.00}{\textbf{#1}}}

\newcommand{\StringTok}[1]{\textcolor[rgb]{0.31,0.60,0.02}{#1}}

\usepackage{hyperref}
\usepackage[utf8]{inputenc}

\usepackage{setspace}
\usepackage{authblk}

\usepackage{amsmath}
\usepackage{graphicx}

\begin{document}

\title{A slice tour for finding hollowness in high-dimensional data}

\author[1,2]{Ursula Laa}
\author[2]{Dianne Cook}
\author[1]{German Valencia}

\affil[1]{School of Physics and Astronomy, Monash University}
\affil[2]{Department of Econometrics and Business Statistics, Monash University}

\maketitle

\begin{abstract}
Taking projections of high-dimensional data is a common analytical and
visualisation technique in statistics for working with high-dimensional
problems. Sectioning, or slicing, through high dimensions is less
common, but can be useful for visualising data with concavities, or
non-linear structure. It is associated with conditional distributions in
statistics, and also linked brushing between plots in interactive data
visualisation. This short technical note describes a simple approach for
slicing in the orthogonal space of projections obtained when running a
tour, thus presenting the viewer with an interpolated sequence of sliced
projections. The method has been implemented in R as an extension to the
tourr package, and can be used to explore for concave and non-linear
structures in multivariate distributions.
\end{abstract}


\hypertarget{introduction}{%
\section{Introduction}\label{introduction}}

Data is commonly high-dimensional, and visualisation often relies on
some form of dimension reduction. This can be done by taking linear
projections, or nonlinear if one considers techniques like
multidimensional scaling (MDS) \citep{mds} or t-Distributed Stochastic
Neighbour Embedding (t-SNE) \citep{tsne}. For the purposes here, the
focus is on linear projections, in particular as provided by the grand
tour \citep{As85, BCAH05}. Interactive and dynamic displays can provide
information beyond what can be achieved in a static display. The grand
tour shows a smooth sequence of interpolated low dimensional
projections, and allows the viewer to extrapolate from the
low-dimensional shapes to the multidimensional distribution. It is
particularly useful for detecting clusters, outliers and non-linear
dependence.

A major limitation of projections is their opacity. It is mitigated
some, when scatterplots are used to render the projected data, through a
pseudo-transparency of sparseness of points. This is improved further if
points are also drawn using an alpha level that provides transparent
dots on most display devices. Some features of multivariate
distributions may be visible, but a lot is easily hidden, especially in
the case of concave structures. Considering the example of a simple
geometric shape such as a hypershere, it is difficult to distinguish
between a full or a hollow sphere, that is, whether points are uniformly
distributed within the sphere or on the surface of a sphere. Similarly
we can think of small scale structures hidden in the centre of a
multivariate distribution, that might be considered to be ``needles in a
haystack'', which can be difficult to detect in projections. Projections
also obscure non-linear boundaries as might be constructed from a
classification model, or nonlinear model fits in high-dimensions.

Slicing, or sectioning, is a way that the internal distribution of
high-dimensional data can be explored. \citet{prosection} discusses a
technique for combining projections with sections constructed by slicing
in the dimensions orthogonal to the projection. It is also possible to
think of linked brushing (e.g.~\citet{ggobiCSDA2003},
\citet{JSSv081i05}) as slicing. In this note, we discuss an approach to
sectioning, where observations in the space orthogonal to the projection
are highlighted if they are close to a projection plane through the mean
of the data, and faded if further afield. This is combined with the
grand tour to provide a new dynamic display that can be used to
systematically search for features hidden in high dimensions.

The new section tour method is described in Section \ref{sec:method},
and it is implemented in the \texttt{tourr} \citep{tourr} package in R
\citep{rref} (Section \ref{sec:implementation}). Section
\ref{sec:examples} illustrates the use on several high-dimensional
geometric shapes and established data sets, showing how concave or
occluded structures can be visualised and explored with a slice tour.
Future work is discussed in Section \ref{sec:discussion}.

\hypertarget{sec:method}{%
\section{Method}\label{sec:method}}

\hypertarget{tour-review}{%
\subsection{Tour review}\label{tour-review}}

A tour provides a continuous sequence of \(d\)-dimensional (typically
\(d=1\) or \(2\)) projections from \(p\)-dimensional Euclidean space. It
is constructed by combining a method for basis selection with geodesic
interpolation between pairs of bases. In a grand tour, the basis
selection is random, each new basis is chosen from all possible
projections. In a guided tour, the bases are chosen based on an index of
interestingness. Different basis selection methods as well as the
geodesic interpolation are implemented in the \texttt{tourr} package
\citep{tourr}, which also provides several display functions for viewing
the tour.

For the explanation of the slice tour, the actual mechanics of the tour
are not important, and only the notion of a projection plane in high
dimensions is needed. The notation used for explanations starts with
denoting each projection as \(\mathbf{X}\) the \(n\times p\) dimensional
data matrix (\(n\) observations in \(p\) dimensions) and \(\mathbf{A}\)
an orthonormal \(p \times d\) projection matrix. The \(d\)-dimensional
projection of the data is thus given by
\(\mathbf{Y} = \mathbf{X}\cdot\mathbf{A}\), producing the \(n\times d\)
dimensional projected data matrix to be plotted in each frame of the
tour display.

To make a section, we are interested in the orthogonal distance of the
data points from a projection plane, particularly using \(d=2\), but the
approach theoretically works for any \(d\).

\hypertarget{slicing-in-the-orthogonal-space}{%
\subsection{Slicing in the orthogonal
space}\label{slicing-in-the-orthogonal-space}}

\hypertarget{distance-from-the-origin}{%
\subsubsection{Distance from the
origin}\label{distance-from-the-origin}}

The orthogonal distance of the data points from the current projection
plane is calculated as,

\begin{equation}
\tilde{v}_i^2 = ||\mathbf{x}_i'||^2,
\end{equation}

\noindent where

\begin{equation}
\mathbf{x}_i' = \mathbf{x}_i - (\mathbf{x}_i\cdot \mathbf{a}_1) \mathbf{a}_1 - (\mathbf{x}_i\cdot \mathbf{a}_2) \mathbf{a}_2
\end{equation}

\noindent and \(\mathbf{x}_i, i=1,...,n\) is a \(p\)-dimensional
observation in \(\mathbf{X}\) and \(\mathbf{a}_k, k=1,2 (=d)\) denoting
the columns of the projection matrix,
\(\mathbf{A}=(\mathbf{a}_1, \mathbf{a}_2)\). \(\mathbf{A}\) is assumed
to be through 0, because the data has been centred on its mean.
\(\mathbf{x}_i'\) can be considered to be a normal from the projection
plane to \(\mathbf{x}_i\), and then the norm of this vector gives the
orthogonal distance between point and plane. This can be generalized for
\(d>2\).

The distance can then be used to display a slice tour by highlighting
points for which the orthogonal distance is smaller than a selected
cutoff value \(h\). In 3D, where for each plane there is only a single
orthogonal direction, this defines a flat slice of height \(2h\). We
could use this single direction to slice systematically from one side of
the space to the other. It is also possible to think of front and back,
some points are closer to the viewer, and some are far, and depth cues
could be used. When going beyond 3D, the dimension of the orthogonal
space is \(>1\), and all sense of direction is lost. In general, there
are \((p-d)\) dimensions orthogonal to the projection plane, and using
distance from the points to the plane is the simplest approach to
generating slices. Using the Euclidean distance results in rotation
invariant slicing in the orthogonal space, where a ``slice'' is
spherical in the orthogonal subspace, and has radius \(h\).

\begin{figure}
\centering
\includegraphics{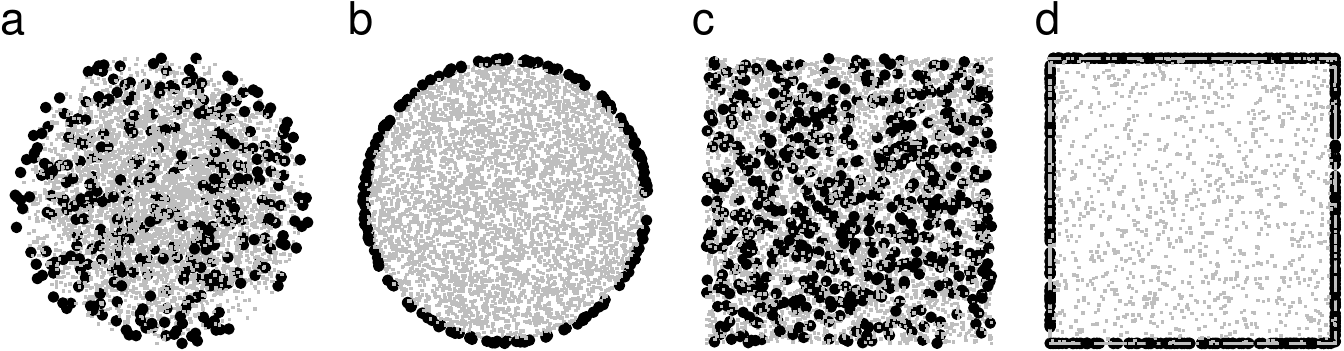}
\caption{Sliced projections through 4D geometric shapes. On the left a
full (hollow) hypersphere (a, b), on the right a full (hollow) hypercube
(c, d). Points inside the slice are shown as black bullets, points
outside the slice are shown as grey dots. We can clearly distinguish the
full from the hollow objects based on the slice
display.\label{fig:renderFunction}}
\end{figure}

\begin{figure*}[ht]
\centerline{\includegraphics[width=0.4\textwidth]{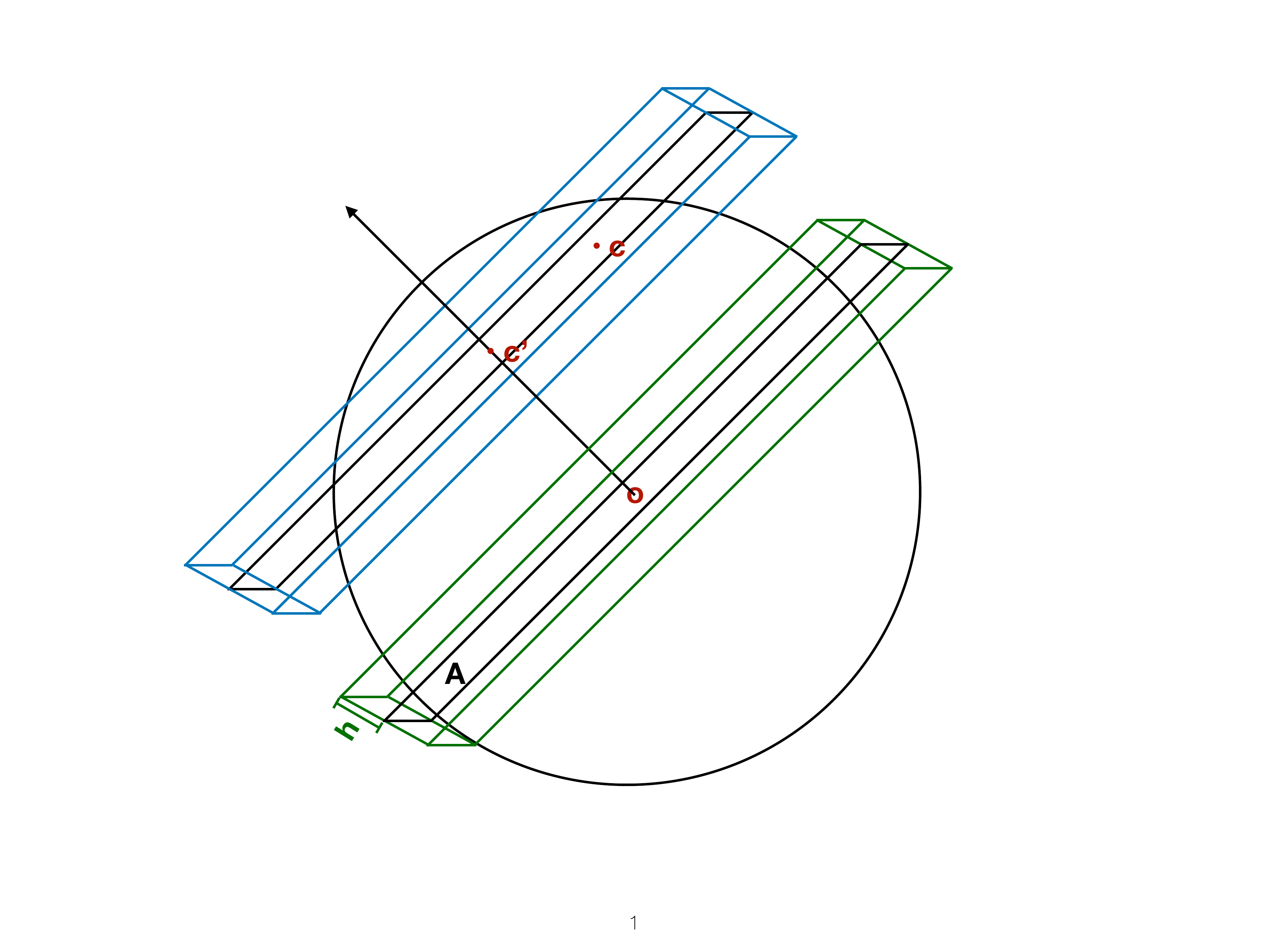}
\hspace{5mm}
\includegraphics[width=0.4\textwidth]{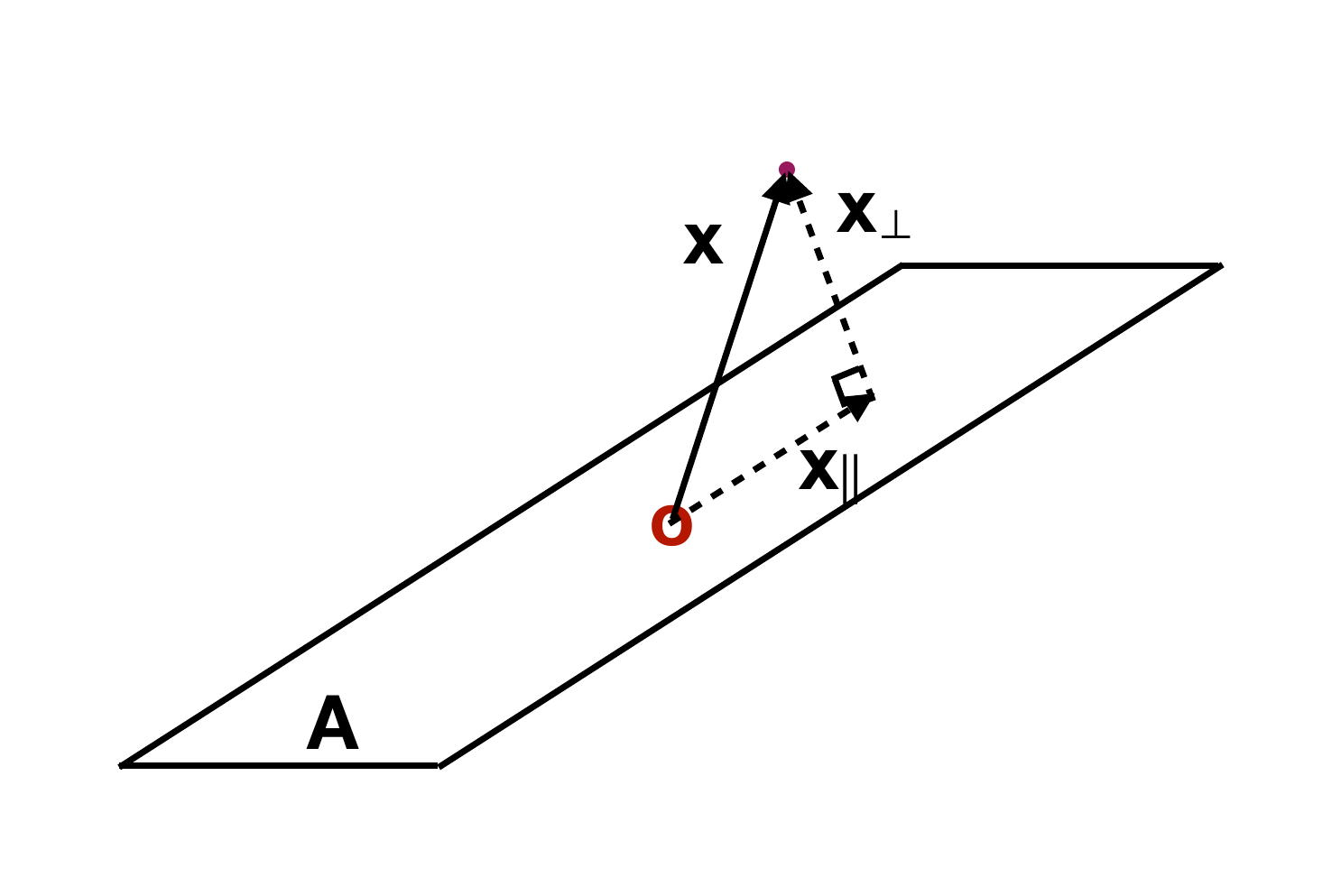}}
\caption{Illustrations of slicing, through a 3D sphere (left), and demonstrating the calculation of the orthogonal distance (right). 3D slicing can be done by sliding the projection plane along the orthogonal direction: centred at the origin (green) and one off-centre at $\mathbf{c}$ (blue). This intuition does not transfer to higher dimensions, and it is best to use orthogonal distance between point and projection plane for computing the slice. Because a projection plane has no specific location, for slicing we can prescribe this, as through the data centre, or any other point, $\mathbf{c}$.}
\label{fig:diagrams}
\end{figure*}

Figure \ref{fig:renderFunction} shows slices through 4D solid and hollow
geometric shapes. For each shape, the points are generated either
uniformly within the shape, or on the surface. This illustrates the ease
of distinguishing the solid from the hollow, once slices are made.
Figure \ref{fig:diagrams} illustrates the slicing method for 3D and
higher dimensions.

\hypertarget{slice-thickness}{%
\subsubsection{Slice thickness}\label{slice-thickness}}

Choosing the slice thickness is a compromise between what feature size
can be resolved and the sparseness of the data. As \(p\) increases, the
relative number of points that are inside a slice of fixed thickness
\(h\) will decrease. The exact relation depends on the distribution of
the data points. To get a best estimate for \(h\), given \(p\), assume
that the points are uniformly distributed in a hypersphere. This is a
rotation invariant uniform distribution in \(p\) space, and because with
slicing we are mostly interested in hollowness this is more relevant
than assuming a multivariate normal distribution.

The fraction of points inside a slice of thickness \(h\) can be
estimated as the relative volume of the slice compared to the volume of
the full hypersphere, \begin{equation}
V_{rel} = \frac{1}{2} \frac{h^{p-2}}{R^p} (p R^2 - (p-2) h^2) \approx \frac{1}{2} (\frac{h}{R})^{p-2},
\end{equation} where \(R\) is the radius of the hypersphere. The
approximation is valid when \(h \ll R\), as is typically expected to be
the case. To keep the relative number of points (i.e.~\(V_{rel}\))
approximately constant for slices in different dimensions \(p\), we
calculate \(h\) from a volume parameter as \(h = \epsilon^{1/(p-2)}\),
where \(\epsilon\) is a pre-chosen value indicating a fraction of the
overall volume to slice, say 0.1.

\hypertarget{non-central-slice}{%
\subsubsection{Non-central slice}\label{non-central-slice}}

The equations above assume that the projection plane, and thus the the
slice, passes through 0, which will be the mean of centred data. This
(centre point) is generally a good option for the slice tour, because as
the tour progresses the projection plane changes and can catch
non-central concavities, too. However it is straightforward to
generalize the equations to use any centre point, \(\mathbf{c}\).

In general \(\mathbf{c}\) can be any point in the \(p\)-dimensional
parameter space, but we are only interested in the orthogonal component,
the part of the vector extending out of the projection plane,
\(\mathbf{c}' = \mathbf{c} - (\mathbf{c}\cdot \mathbf{a}_1) \mathbf{a}_1 - (\mathbf{c}\cdot \mathbf{a}_2 )\mathbf{a}_2\).
The generalized measure of orthogonal distance is then \begin{equation}
v_i^2 = ||\mathbf{x}_i' - \mathbf{c}'||^2 = \mathbf{x}_i'^2 + \mathbf{c}'^2 - 2 \mathbf{x}_i'\cdot\mathbf{c}'
\end{equation} where the cross term can be expressed as \begin{equation}
\mathbf{x}_i'\cdot\mathbf{c}' = \mathbf{x}_i\cdot\mathbf{c} - (\mathbf{c}\cdot \mathbf{a}_1) (\mathbf{x}_i\cdot \mathbf{a}_1) - (\mathbf{c}\cdot \mathbf{a}_2) (\mathbf{x}_i\cdot \mathbf{a}_2).
\end{equation}

Using the generalized distance measure with a cutoff volume,
\(\epsilon\), then corresponds to moving a slice of fixed thickness,
corresponding to a neighbourhood of the projection plane through the
centre point \(\mathbf{c}\). In 3D, this simply corresponds to moving up
or down along the orthogonal direction. Note that moving \(\mathbf{c}\)
off-centre will result in fewer points inside a slice for most data.

\hypertarget{sec:implementation}{%
\section{Implementation}\label{sec:implementation}}

The slice tour has been implemented in R \citep{rref} as a new display
method \texttt{display\_slice} in the \texttt{tourr} package
\citep{tourr}. In addition to usual parameters the user can choose the
volume parameter \(\epsilon\) by setting \texttt{eps} and, if required,
the centre point \(\mathbf{c}\) is set by the \texttt{anchor} argument.
By default \texttt{eps=0.1} and \texttt{anchor=NULL}, resulting in
slicing through the mean. In addition the user can select the marker
symbols for point inside and outside the slice. By default points in the
slice are highlighted as \texttt{pch\_slice\ =\ 20} and
\texttt{pch\_other\ =\ 46}, i.e.~plotting a bullet for points inside the
slice, and a dot for points outside. Below we show example code for
displaying slices through a hollow 3D sphere.

\begin{Shaded}
\begin{Highlighting}[]
\KeywordTok{library}\NormalTok{(tourr)}
\CommentTok{# use geozoo to generate points on a hollow 3D sphere}
\NormalTok{sphere3 <-}\StringTok{ }\NormalTok{geozoo}\OperatorTok{::}\KeywordTok{sphere.hollow}\NormalTok{(}\DecValTok{3}\NormalTok{)}\OperatorTok{$}\NormalTok{points}
\KeywordTok{colnames}\NormalTok{(sphere3) <-}\StringTok{ }\KeywordTok{c}\NormalTok{(}\StringTok{"x1"}\NormalTok{, }\StringTok{"x2"}\NormalTok{, }\StringTok{"x3"}\NormalTok{) }\CommentTok{# naming variables}
\CommentTok{# slice tour animation with default settings}
\KeywordTok{animate_slice}\NormalTok{(sphere3)}
\CommentTok{# trying an off-center anchor point, thicker slice, and}
\CommentTok{# we use pch=26 to hide points outside the slice}
\NormalTok{anchor3 <-}\StringTok{ }\KeywordTok{rep}\NormalTok{(}\FloatTok{0.7}\NormalTok{, }\DecValTok{3}\NormalTok{)}
\KeywordTok{animate_slice}\NormalTok{(sphere3, }\DataTypeTok{anchor =}\NormalTok{ anchor3, }\DataTypeTok{eps =} \FloatTok{0.2}\NormalTok{, }\DataTypeTok{pch_other=}\DecValTok{26}\NormalTok{)}
\end{Highlighting}
\end{Shaded}

\hypertarget{sec:examples}{%
\section{Examples}\label{sec:examples}}

\hypertarget{geometric-shapes}{%
\subsection{Geometric shapes}\label{geometric-shapes}}

Using the \texttt{geozoo} package \citep{geozoo} a number of ideal
shapes are generated.

\hypertarget{hollow-sphere}{%
\subsubsection{Hollow sphere}\label{hollow-sphere}}

We sample points on the surface of a sphere with radius \(R=1\) in
\(p=3\) and \(p=5\) dimensions. Using the \texttt{sphere.hollow()}
function, we generate two sample data sets by generating 2000 and 5000
points from a uniform distribution on the surface of a 3D and 5D
spheres. For both examples we start by slicing through the origin with
the default parameters, in particular \(\epsilon = 0.1\),
i.e.~\(h = 0.1\) in 3D and \(h = 0.46\) in 5D. Slicing through the
origin results in sections where the points inside the slice are
approximately on a circle with the full radius, and each view is similar
to that shown in Figure \ref{fig:renderFunction}.

\begin{figure*}[ht]
\centerline{\includegraphics[width=0.2\textwidth]{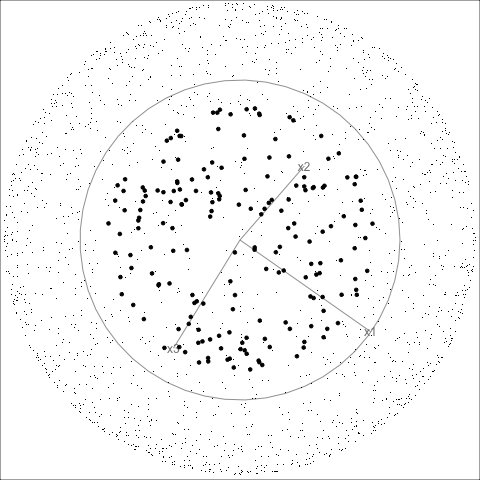}
\hspace{5mm}
\includegraphics[width=0.2\textwidth]{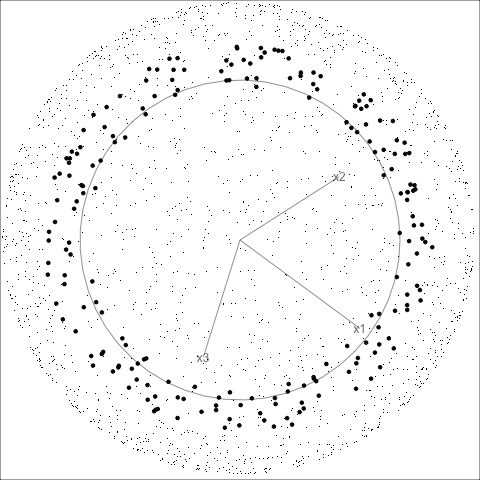}
\hspace{5mm}
\includegraphics[width=0.2\textwidth]{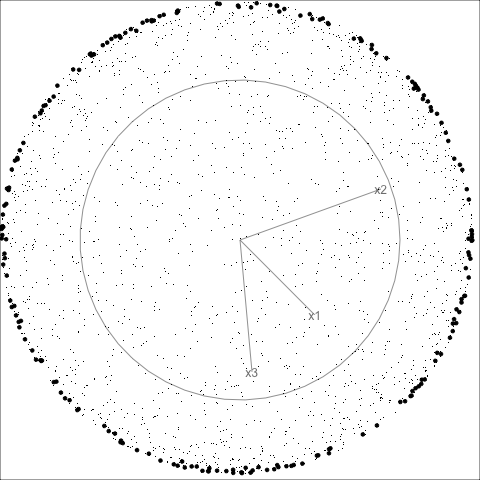}}
\centerline{\includegraphics[width=0.2\textwidth]{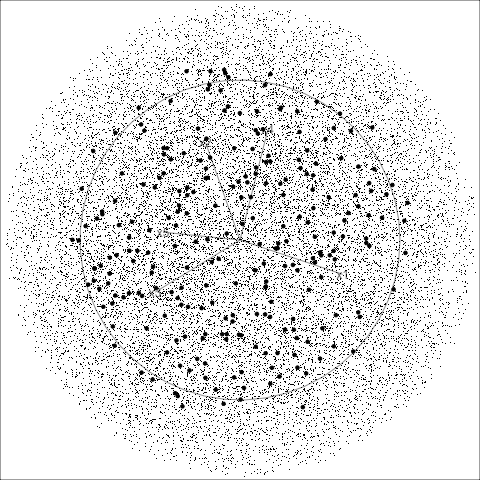}
\hspace{5mm}
\includegraphics[width=0.2\textwidth]{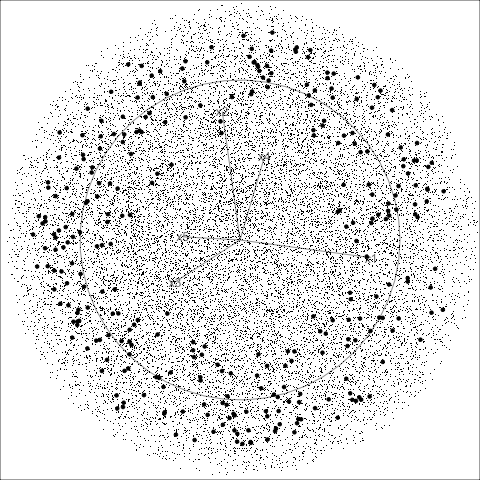}
\hspace{5mm}
\includegraphics[width=0.2\textwidth]{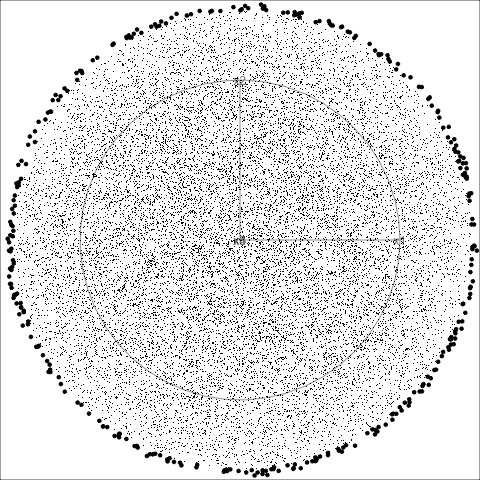}}
\caption{Different slices through a 3D (first row) and 5D (second row) hollow sphere with $R=1$, with shifted centre point, showing full circle with small radius (left), circle with radius $<R$ (middle) and circle with radius $R$ (right).}
\label{fig:3anchored}
\end{figure*}

It is especially instructive to look at spheres that are sliced
off-centre. In this case the different views obtained in the slice tour
reveal more of the concave structure of the distribution. The first row
of Figure \ref{fig:3anchored} shows example views from a slice tour on a
3D sphere with \(R=1\) and centre point \((0.7, 0.7, 0.7)\). Notice that
this was chosen to fall outside the sphere. Depending on the viewing
angle selected by the tour the slice contains points on the circle with
radius \(R\) (when the centre point has a negligible orthogonal
component to the viewing plane); a circle with radius \(<R\) as the
viewing angle is tilted away from the axis connecting the centre point
to the origin; a full circle with small radius as the angle increases;
and finally we see an empty slice when the projection plane is
orthogonal to the centre point axis. A similar picture is found for the
5D sphere, see second row of Figure \ref{fig:3anchored}. For this
example we generate 20k points on a hollow 5D sphere to resolve the
features. Since \(h\) increases with \(p\) the resolution is reduced
compared to the 3D example. Note also that as dimensionality increases,
the larger orthogonal space means that the centre point will have a
large orthogonal component in most views.

\hypertarget{other-geometric-shapes}{%
\subsubsection{Other geometric shapes}\label{other-geometric-shapes}}

\begin{figure*}[ht]
\centerline{\includegraphics[width=0.2\textwidth]{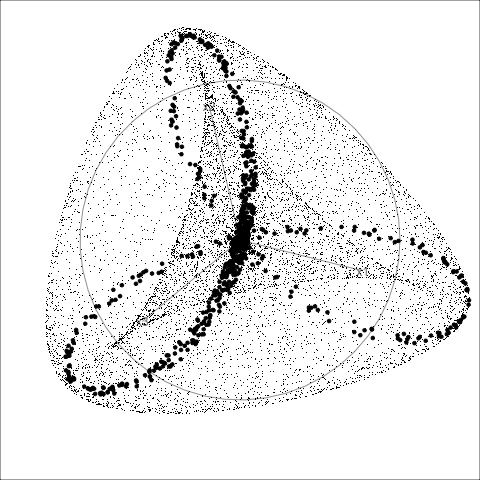}
\hspace{5mm}
\includegraphics[width=0.2\textwidth]{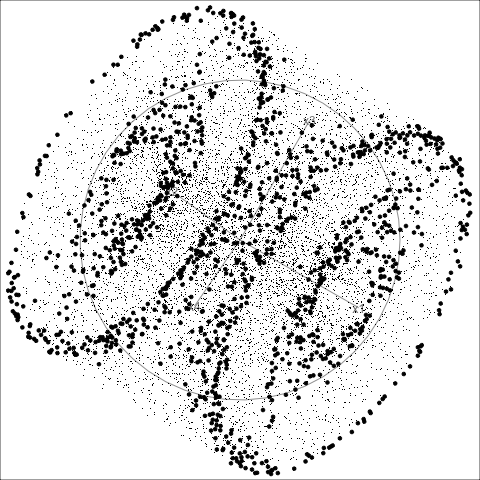}
\hspace{5mm}
\includegraphics[width=0.2\textwidth]{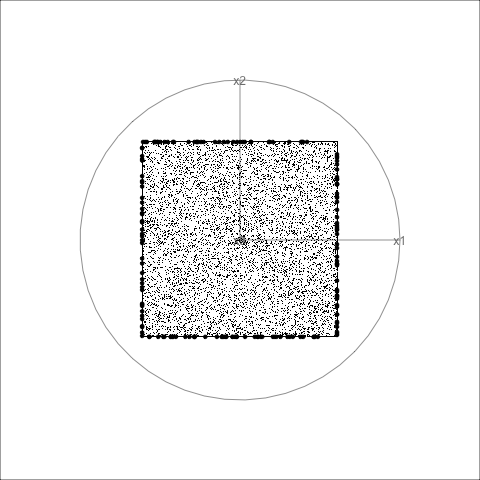}}
\centerline{
\includegraphics[width=0.2\textwidth]{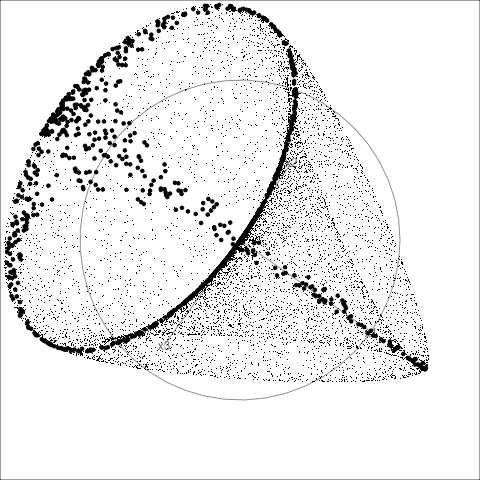}
\hspace{5mm}
\includegraphics[width=0.2\textwidth]{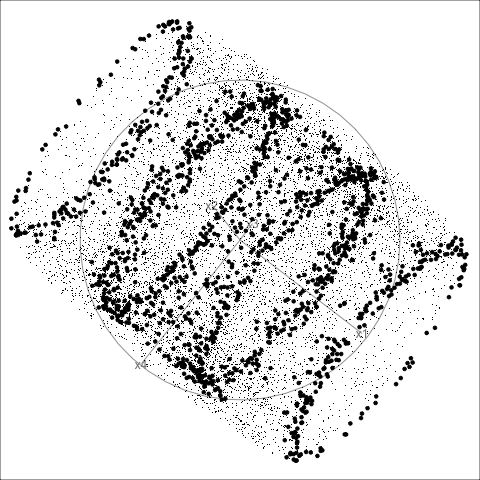}
\hspace{5mm}
\includegraphics[width=0.2\textwidth]{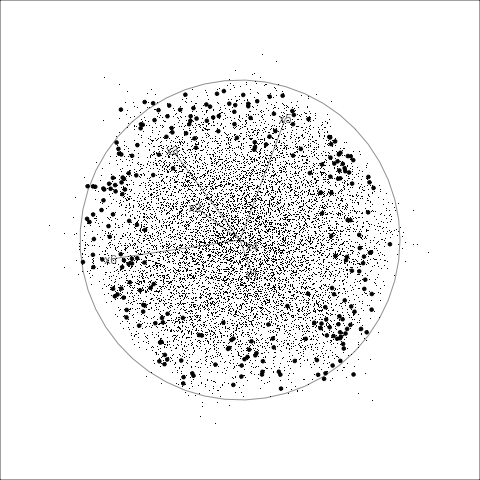}
}
\caption{Different slices through a Roman surface (left), 4D torus (middle) and a 6d cube (right).}
\label{fig:geoms}
\end{figure*}

To better understand the slice tour we look at different examples of
geometric shapes, see Figure \ref{fig:geoms}. For each shape two
selected views are shown. The first column shows views from the slice
tour on a 3D Roman surface. The second column shows slices through a 4D
torus, revealing different aspects of the shape. The last column shows a
6d cube, where the upper plot shows a view along two of the original
parameters, allowing to clearly identify the rectangular shape. The
panel below shows that this is not typically the case when looking at a
randomly selected slice.

\hypertarget{other-examples}{%
\subsection{Other examples}\label{other-examples}}

\hypertarget{needle-in-a-haystack}{%
\subsubsection{Needle in a haystack}\label{needle-in-a-haystack}}

\begin{figure*}[ht]
\centerline{\includegraphics[width=0.2\textwidth]{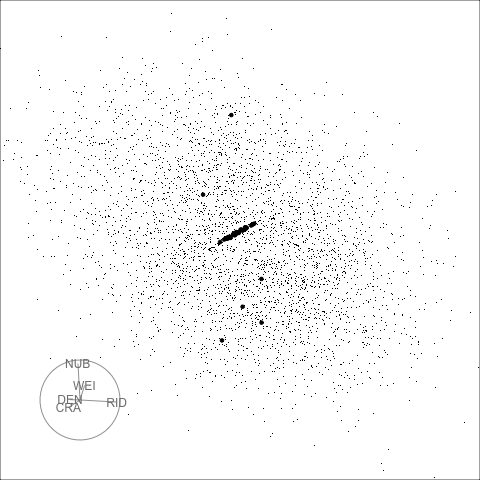}
\hspace{5mm}
\includegraphics[width=0.2\textwidth]{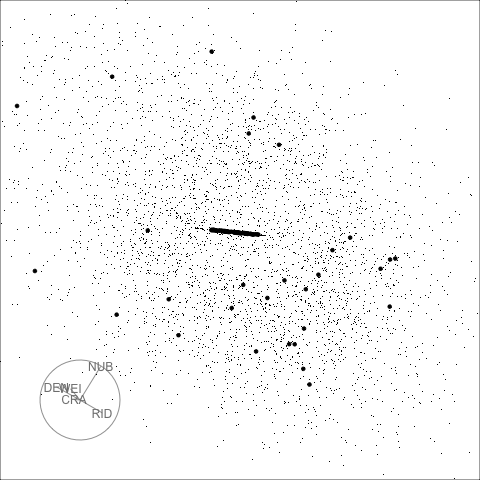}
\hspace{5mm}
\includegraphics[width=0.2\textwidth]{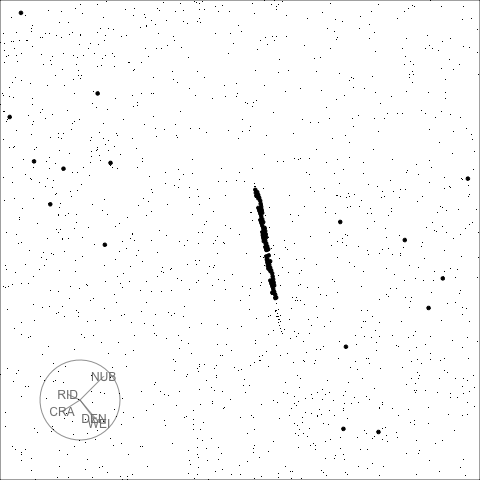}
\hspace{5mm}
\includegraphics[width=0.2\textwidth]{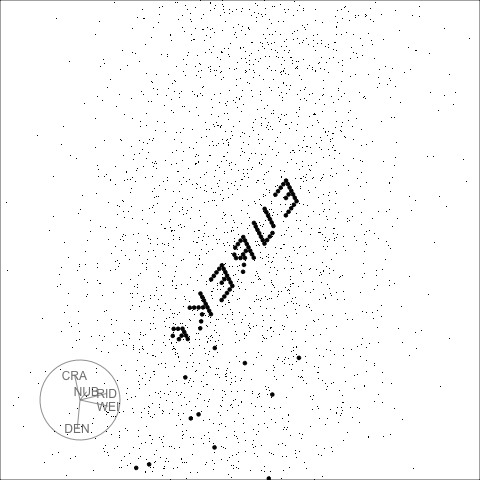}
}
\caption{Different slices through the classic pollen dataset. The first two plots have slice volume $\epsilon = 0.0005$, the last two plots zoom in on the centre and have increased volume $\epsilon = 0.005$. The hidden word can be seen in the zoomed slice.}
\label{fig:pollen}
\end{figure*}

We use the pollen data as an example for a hidden feature generally
occluded in projections. This is a classic 5D data set, originally
simulated by David Coleman of RCA Labs, for the Joint Statistics
Meetings 1986 Data Expo \citep{pollen}. The standardised data is
observed in a slice tour with a thin slice (\(\epsilon = 0.0005\)).
Selected views are shown in the first two plots in Figure
\ref{fig:pollen}. They indicate the presence of an interesting feature
hidden in the centre, which can be identified as the word ``EUREKA'' by
zooming.

\hypertarget{non-linear-boundaries}{%
\subsubsection{Non-linear boundaries}\label{non-linear-boundaries}}

\citet{sam.11271} described some principles and approaches for
visualising models in the data space. Much of this is based on examining
projections provided by a tour of the model in the high-dimensional
space. Classification boundaries are explored for the wine data set
\citep{asuncion:2007}. It is difficult to digest the boundaries fully --
for example, where one group's boundary wraps another, if they are
linear or nonlinear, or whether the boundary goes through the space or
is only carving out a corner of it. The sliced projections makes this
easier.

\begin{figure*}[ht]
\centerline{
\includegraphics[width=0.3\textwidth]{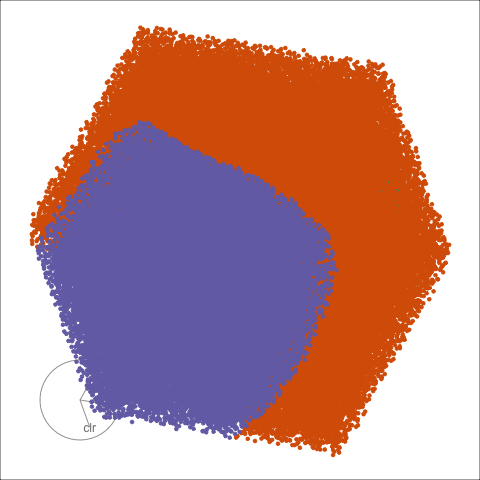}
\hspace{5mm}
\includegraphics[width=0.3\textwidth]{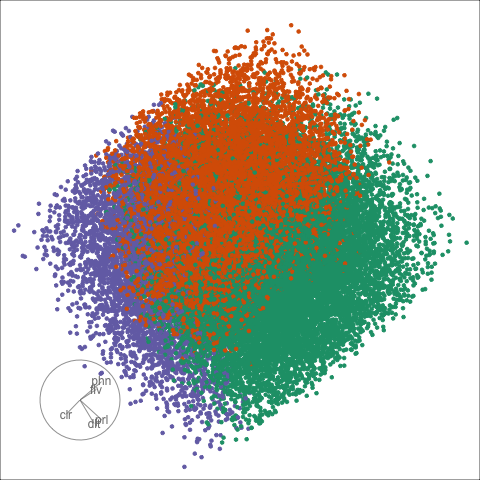}
\includegraphics[width=0.3\textwidth]{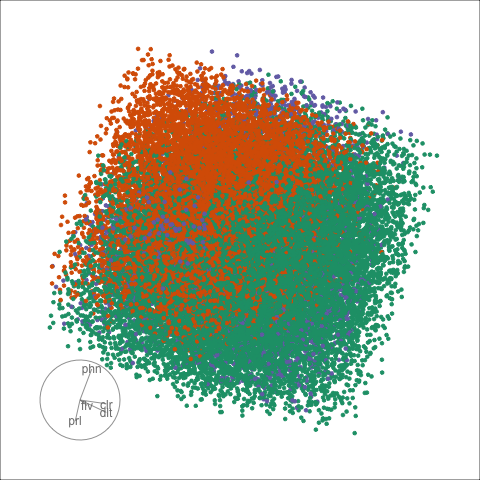}}
\centerline{
\includegraphics[width=0.3\textwidth]{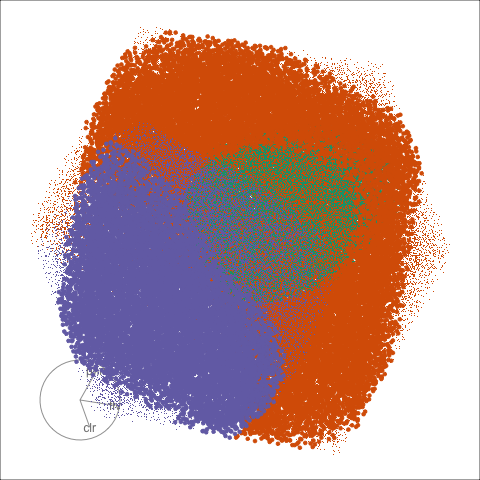}
\hspace{5mm}
\includegraphics[width=0.3\textwidth]{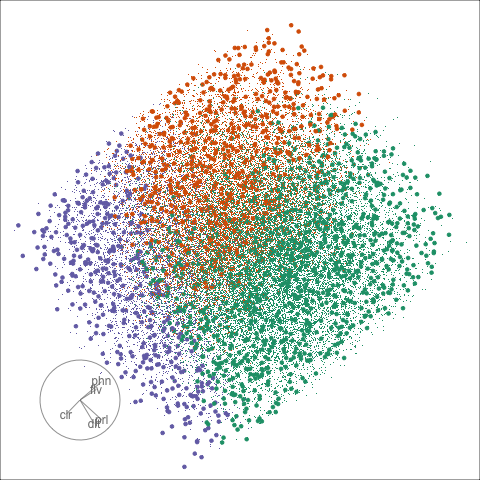}
\includegraphics[width=0.3\textwidth]{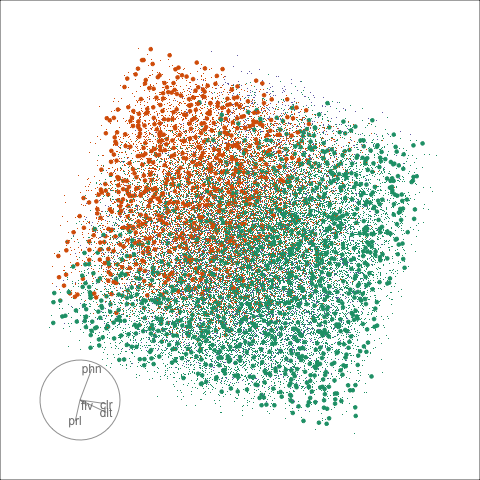}
}
\caption{Exploring classification boundaries of the wine dataset using projection (top row) and sliced projection (bottom row). Radial basis SVM (first column) of 3 variables shows how the slicing reveals the spherical shape of the boundary of one group (green) that was hidden in the projection. Polynomial basis SVM (second and third columns) of 5 variables. The orange group does tend to be wrapped by the green group, and the blue group disappears with slicing, showing that it is on the outer edge of the space.}
\label{fig:wine}
\end{figure*}

Selected views of projections and sliced projections from a radial basis
SVM on 3 variables, and a polynomial basis SVM on 5 variables are shown
in Figure \ref{fig:wine}. The slicing allows exploring the centre of the
space. With 3D it reveals the spherical boundary of the group (green)
hidden by the projection. In 5D with the polynomial basis, projection
might suggest that the boundary between classes is almost linear, but
the slicing shows that it to be nonlinear near the centre.

\hypertarget{sec:discussion}{%
\section{Discussion}\label{sec:discussion}}

This paper has introduced a new visualization method for dynamic slicing
of high-dimensional spaces. It is based on interpolated projections
obtained in a (grand) tour and generates an interpolated sequence of
sliced projections. The examples shown in Section \ref{sec:examples}
demonstrate the potential of this new display to find and explore
concave structures, as well as other hidden features.

A default slice thickness is provided with the algorithm, that takes
dimensionality into account. As the data dimension increases, more
points in the sample are needed, and a thicker slice may be needed.
Generally, the tour can be slow to view when there are a large number of
samples, and the slicing, where only points inside the slice are drawn,
might also be a way to improve the display drawing, with a focus on the
important features.

Sliced 2D tours were available in XGobi \citep{SCB98} but were not
documented. This work makes them available in the \texttt{tourr}
package. It is a simple, but effective, approach to taking slices.

The approach by \citet{prosection} is more complex, and slices a
subspace of the \(p-d=p-2\) dimensional space orthogonal to the
projection. This generates a more parameters, making it more difficult
to navigate. More parameters mean more decisions on what to show.
However, it is one of the next steps to explore different definitions of
a slice tour, and the types of structure that might be captured by
variations in slicing algorithms.

Lastly, there is a large literature on projection pursuit, and some of
this work is available in the projection pursuit guided tour in the
\texttt{tourr} package. The projections shown are more interesting in a
guided tour than might be seen using a grand tour. The slicing might be
introduced into projection pursuit by defining weighted projection
pursuit indexes. The resulting indexes could be incorporated into a
guided slice tour, finding projections where the slice reveals something
new.

\hypertarget{acknowledgements}{%
\section{Acknowledgements}\label{acknowledgements}}

The authors gratefully acknowledge the support of the Australian
Research Council. The paper was written in \texttt{rmarkdown}
\citep{rmarkdown} using \texttt{knitr} \citep{knitr}. The source
material and animated gifs for this paper are available at
\url{https://github.com/uschiLaa/paper-slice-tour}. An appendix deriving
the relative slice volume for the hypersphere is included as
supplemental material.

\bibliographystyle{tfcad}
\bibliography{biblio.bib}

\end{document}